\documentclass[journal=jacsat,manuscript=article]{achemso}

\usepackage[version=3]{mhchem} 
\usepackage{graphicx}
\usepackage{dcolumn}
\usepackage{bm}
\usepackage{natbib}
\usepackage{amssymb}

\author{J. C. Garcia}
\affiliation[Escola Polit\'ecnica, Universidade de S\~ao Paulo,
CP 61548, CEP 05424-970, S\~ao Paulo, SP, Brazil]
{Universidade de S\~ao Paulo, S\~ao Paulo, SP, Brazil}
{}
\author{L. V. C. Assali}
\affiliation[Instituto de F\'{\i}sica,Universidade de S\~ao Paulo,
CP 66318, CEP 05315-970, S\~ao Paulo, SP, Brazil]
{}
\author{J. F. Justo}
\email{jjusto@lme.usp.br}
\affiliation[Escola Polit\'ecnica, Universidade de S\~ao Paulo,
CP 61548, CEP 05424-970, S\~ao Paulo, SP, Brazil]
{Universidade de S\~ao Paulo, S\~ao Paulo, SP, Brazil}

\title[nanowire]
{The structural and electronic properties of tin oxide nanowires:
an {\it ab initio} investigation}

\begin{document}

\begin{abstract}
We performed an {\it ab initio} investigation on the properties of
rutile tin oxide (SnO$_{x}$) nanowires.
We computed the wire properties determining the equilibrium geometries,
binding energies and electronic band structures for several wire
dimensions and surface facet configurations.
The results allowed to establish scaling laws for the
structural properties, in terms of the nanowire perimeters.
The results also showed that
the surface states control most of the electronic properties of
the nanowires. Oxygen incorporation in the nanowire surfaces
passivated the surface-related electronic states, and the
resulting quantum properties and
scaling laws were fully consistent with electrons confined inside
the nanowire. Additionally, oxygen incorporation in the wire
surfaces generated an unbalanced concentration of spin up and down electrons,
leading to magnetic states for the nanowires.
\end{abstract}


\section{Introduction}

Over the last decade, there has been growing interest in
semiconducting one-dimensional
nanostructures \cite{cui,comini2009}. They open
possibilities for quantum confinement, which may allow to obtain
tailored electronic properties, such as optical transitions in
pre-determined wavelengths and selective electronic response from
interaction with specific
molecules \cite{lu2006}. Among many semiconducting nanostructured
materials, tin oxide (SnO$_{x}$) nanowires have received special
attention, mostly due to several promising applications, such as gas
\cite{Kolmakov2003,Kolmakov2005}, chemical \cite{Sysoev2007}
and humidity \cite{Kuang2007}  sensors, solar cells \cite{Gubbala2008},
optical devices \cite{Luo2006,Chen2009a},
and high-density batteries \cite{Park2007,Zhong2011}. For sensors,
nanostructured tin oxide represents an improvement over the previous
devices based on bulk tin oxide, since they provide greater
surface/volume ratio, allowing to get higher sensitivity and
selectivity for several chemical species.

Tin oxide nanowires have been grown, in rutile configuration,
along the [001] \cite{Qin2008a}, [011] \cite{Zhong2011},
[101] \cite{Dai2002,Lupan2009,Kar2011,Ph2011a},  and
[121] \cite{Ph2011a} directions. Nanostructured tin oxide have
also been recently grown in other forms, such as
nanotubes \cite{Dai2002}, nanoribbons \cite{Hu2002}, and
nanorods \cite{Huang2011}. Those one-dimensional nanostructures
have been synthesized by several processes, such as
wet-chemical approach \cite{Qin2008a}, high temperature thermal
oxide  \cite{ Dai2002}, hydrothermal \cite{Lupan2009} and
vapor-liquid-solid \cite{Kar2011} methods, carbothermal
reduction \cite{Ph2011a},  plasma-enhanced chemical vapor
deposition \cite{Huang2011}, and through oxidation of tin vapors
at elevated temperatures \cite{Hu2002}.

Square-shaped rutile tin oxide nanowires, along the [001] direction,
have been obtained with diameters of around 80 nm and lengths
of a few micrometers \cite{Qin2008a}. Tin oxide nanowires with
rectangular cross sections have also been grown along the [101]
direction, leading to structures with diameters of 50-150 nm and
lengths of around 10-100 $\mu$m. Those results indicate very large
length/diameter ratios, which may be important for incorporation
in integrated circuits.

There are several questions that still need to be addressed to
optimize the use of tin oxide nanowires as
nanosensors \cite{Kolmakov2004}. For example, several sensing
properties require attention, such as the response time, sensitivity,
selectivity, and degradation to long-term exposure to gases.
In order to optimize the sensing performance, it is important to
understand the fundamental properties of those tin oxide nanowires,
and how those properties scale. Here, we explored the electronic
and structural properties of rutile [100] SnO$_x$ nanowires,
with several facet configurations, using simulations based
on first-principles total energy methodologies. We observed
that the scaling laws of the structural properties could be described
in terms of the amount of nanowire surface, independent of the wire
shape, which is expressed in terms of the nanowire perimeter,
consistent with what was observed for nanowires of different
materials \cite{Justo2007a,Justo2007b}. We also showed that
the electronic properties of the nanowires are strongly affected
by the oxygen passivation of surface states.
Additionally, we observed that such passivations lead to
magnetic states for the nanowires.

\section{Methodology}

Although the sensing properties and the growth procedures of
tin oxide nanowires have been intensively studied, there is
scarce literature that covers the theoretical modeling of this
material. The electronic quantum confinement in tin oxide nanowires,
with artificial surface passivation, have been investigated
within the density functional theory \cite{Deng2010}. That investigation
showed
that the wire bandgap scales with the inverse of the wire diameter,
since the surface-related energy levels were fully removed by
the surface passivation. On the other hand, there are similarities
on the structural properties of tin oxide and titanium oxide
nanowires, such that their properties are general compared by
theoretical investigations. In the case of titanium oxide
nanowires, there is a more extensive literature, that covers the
theoretical investigations on their structural and electronic
properties \cite{Tafen2009,Cakir2009,Migas2010,He2011,Aradi2011}.

Our calculations on tin oxide nanowires were performed using
the Vienna {\it ab initio} simulation package (VASP) \cite{kresse1}.
The electronic exchange-correlation potential was described within
the spin-polarized density functional theory and the generalized
gradient approximation (DFT-GGA) \cite{pbe}. The electronic
wave-functions were described by a projector augmented wave (PAW)
method \cite{kresse2}, taking a plane-wave basis set with a kinetic
energy cutoff of 400 eV. Self-consistent calculations were performed
until reaching convergence in total energy of 1 meV between two
consecutive iterations. Configurational optimization was performed
by considering relaxation in all atoms, without any symmetry
constrain, until forces were smaller than 3 meV/\AA~ in any atom.
The Brillouin zone was sampled by a $1\times 1 \times 11$ k-point
grid \cite{mp}. The structures were built using periodic boundary
conditions with a tetragonal simulation cell. In the directions
perpendicular to the nanowire one ($z$), lattice parameters were
chosen such that there was a large open space between the atoms
in the original cell and those in the image ones. We found that
an open space of about 15 \AA~ in any direction was large enough
to prevent interactions between the atoms in the simulation cell
with those in the neighboring image cells.

\section{Results}

Tin dioxide (SnO$_{2}$) in a crystalline bulk phase has a
tetragonal structure with the space group
$\mathop D\nolimits_{4h}^{14}$ (P4/mnm, 136), with
experimental \cite{Hazen} (calculated) lattice
parameters $a_{bulk}= 4.737\, (4.832)$ \, \AA\,
$c_{bulk}=3.186\, (3.247)$ \, \AA, $u=0.3064\, (0.3065)$,
resulting in Sn-O average interatomic distance of
2.054 (2.094) \, \AA.
Bulk tin oxide is a wide band gap (E$_{g}$=3.6 eV) metal oxide
semiconductor, as measured by optical absorption \cite{Agekyan}.
Our calculations for the bulk crystal gave a value
of 0.65 eV for the direct band gap at the $\Gamma$ point. Our results
in terms of interatomic distances and bandgap were
in good agreement with other theoretical investigations
based on the density functional theory \cite{Borges}.
Therefore, the present investigation also observed the general
trend of DFT calculations to underestimate
the bandgap of bulk SnO$_{2}$ and
DFT-GGA to overestimate the lattice parameters \cite{Varley}.

Tin oxide (SnO$_{x}$)
nanowires have been observed to grow in the rutile configuration,
in several growing directions, but most of the investigations
are associated with wires along the [001] direction.
Additionally, due to the strong interatomic Sn-O interactions,
those nanowires have a core that resembles the structure of their
crystalline counterpart. Therefore, we focused our investigation
on the properties of nanowires with a crystalline core grown along
the [001] direction, and several facet configurations.

\ref{fig1} presents the cross section of the tin oxide nanowires
investigated here. We initially considered the properties of
nanowires without any surface passivation, only later we observed
the role of oxygen passivation on the electronic properties of the wires.
The figure represents the relaxed final configurations of the
simulations for the nanowires with pure \{110\} surface facets
(R9, R16, R25, R36, and R49)
and with a mixture of \{100\} and \{110\} ones
(R21, R37, R45, and R69). The results indicated
that tin oxide nanowires keep their rutile-like structure even for
the smallest dimensions, such as the nanowires with labels R9 and R21,
with both types of surface configurations.

Determining scaling laws for the properties of nanowires has been a
challenging task, mainly when trying to compare the properties of wires
with different surface facets, or even with different growing directions.
Several attempts to establish those scaling laws, in terms of the wire
cross sections, diameter, or even density of atoms in the wire, have failed.
The diameter is generally chosen as the dimensional parameter in which
a scaling law is built \cite{rurali,kizuka}. However, defining a unique
nanowire diameter is not simple, since those nanowires generally have
facets and do not have a single diameter. Authors either avoid defining
such a parameter \cite{pono,rurali}, or describe the wire representative
dimension as the smallest wire diameter, taken from images of the wire cross
section \cite{kizuka}.  Ultimately, it is generally assumed that the
nanowires have a prevailing cylindrical shape \cite{kagimura,gulseren}.
For nanowires with large diameters ($>$ 5nm), properties are generally well
described using any of those assumptions, but for thinner wires those
models clearly fail. As a result, the scaling laws in terms of those
dimensional physical parameters are valid only within a specific wire family,
in which all the wires have the same growth direction and surface types.

The literature lacks a unified model that could put together nanowires of
a certain material with all types of surfaces, facets, and growth directions.
Recently, interatomic potentials \cite{edip97,edip98} have been used
to show that the nanowire scaling laws could be well described
in terms of the respective wire perimeters \cite{Justo2007a, Justo2007b}.
The relevance of such dimensional parameter was not casual, the wire
perimeter ({\rm P}), the sum of all sides of the wire cross section,
multiplied by the length of the wire ({\rm L}) gives
the total nanowire surface area ({\rm S = P$ \times$ L}). It is well
known that the ratio of surface/volume of nanowires is very large, such
that the properties of those nanostructures would scale with the amount
of nanowire surface (({\rm S}). \ref{fig4} shows the binding energy of all
the nanowires presented in \ref{fig1}.
The figure presents the energy as a function of the inverse perimeter
({\rm 1/P}) of the nanowire. It gets clear that each family of wire
type has an specific trend in energy, scaling with the inverse perimeter,
all going to the same value for the bulk (as the perimeter tends to infinite).

The results showed that, for each family, there is an almost linear relation
between binding energy and {\rm 1/P}, for a wide range of wire perimeters.
The trend only deviates from a linear behavior for very small
nanowires (large {\rm 1/P}), for example beyond the R16 wire.
This is a reasonable result, considering that for those very thin nanowires,
the strong rutile-like structure starts to weaken its rigidity and the systems
relax toward more favorable configurations.

Table I summarizes the structural properties of all nanowires
studied here. The Sn-O average interatomic distances, for thin
nanowires, are larger than the value for the three-dimensional bulk
SnO$_2$ crystal. On the other hand,
the Sn-O average interatomic distances, of those atoms sitting in the
nanowire surface, remain much larger than the bulk value for all wires.

\ref{fig2} shows the electronic band structure of nanowires of the two
families (R49 and R69), and their respective
states near the top of the valence band and in the bottom of conduction band.
It gets clear that the the surface states control the
properties of the bandgap for excitations of both electrons
and holes. \ref{fig3} presents the effects of
incorporating oxygen atoms in the surfaces of one of the nanowires (R9).
First, for the unpassivated nanowire, there is a large number of
states that stay in the nanowire bandgap, as observed in \ref{fig2}.
As oxygen is introduced in the surface, those states are removed from
the bandgap, as those states move toward the valence and conduction bands.
Additionally, the incorporation of oxygen atoms in the surface
generates an unbalanced distribution between up and down states,
leading to a magnetic state. This result appears very appealing, since
those wires could be used as devices based on spin polarized states.

The oxygen passivation could be better understood in terms of the
modifications in the electronic density of states upon passivation.
\ref{fig3a} presents the density of stats for the R9 nanowire
without and with oxygen passivation. Two main effects were observed.
First, the oxygen passivation essentially moved the conduction
bands upward, opening a bandgap. Second, it lead to a modification of
the highest occupied energy levels, with important oxygen p-related levels
in the top of the valence band. Those oxygen p-related levels are responsible
for the resulting magnetic effects in the nanowires.

\section{Summary}

In summary, we carried a theoretical investigation on the structural and electronic
properties of tin oxide nanowires. We found that the scaling laws of
structural properties could be well described in terms of the nanowire perimeter.
We also observed that the nanowires kept their rutile-like configurations
even for the thinnest wires. In terms of the electronic structures,
we found that the surface states control the bandgap states for unpassivated
nanowires. Those surface-related states were fully passivated upon oxygen
incorporation. The resulting nanowires presented non negligible spin polarization,
coming from the p-related stated of the surface oxygen atoms.
\vspace{1cm}

\noindent
{\bf Acknowledgments:}
The authors acknowledge partial support from Brazilian agencies
 FAPESP and CNPq.

\pagebreak

\begin{table}[!h]
\begin{center}
\caption{Structural properties of tin oxide nanowires. The table presents
the number of Sn and O atoms in the simulation cell, the binding energy
(E$_{b}$) per number of atoms in the unit cell (E$_{b}$/atom), the perimeter (P),
the inverse of perimeter (1/P) and diameter (D) of nanowires. The table also
presents the average interatomic distances Sn-O in the nanowire and in the surface.
Energies are given in eV and  distances in \AA. The  d$_{surf}$(Sn-O) values are
around {\rm 1.6\%} larger than the theoretical values of the bulk SnO$_{2}$.}
\vspace*{0.5cm}
\begin{tabular}{ccccccccc}\hline \hline
NW   & \multicolumn{2}{c}{atom/cell}& E$_{b}/atom$ & P & 1/P & D &
d$_{NW}$(Sn-O)&d$_{surf}$(Sn-O)  \\
   & Sn  & O                      & (eV)         & (\AA)    &
(\AA)$^{-1}$  & (\AA)   & (\AA)&(\AA) \\ \hline
R9   &9 &       12                      &-4.37        &28.1      &0.036    &9.9        &2.109 &2.128 \\
R16  &16&       24                          &-4.46        &42.6      &0.024    &15.1   &2.104 &2.116 \\
R25  &25&       40                          &-4.50        &56.1      &0.018    &19.8   &2.098 &2.124 \\
R36  &36&       60                          &-4.53        &69.1      &0.015    &24.4   &2.097 &2.126 \\
R49  &49&       84                          &-4.56        &84.2      &0.012    &29.8   &2.102 &2.143 \\
\hline
R21  &21&       32                          &-4.53        &47.0      &0.021    & 14.7  &2.112 & 2.142 \\
R37  &37&       60                          &-4.59        &66.5      &0.015    & 19.4  &2.105 & 2.131 \\
R45  &45&       76                          &-4.59        &65.6      &0.015    & 24.4  &2.100 & 2.132 \\
R69  &69&       120                         &-4.65        &93.9      &0.011    & 29.0  &2.096 & 2.116 \\
\hline
bulk &2 & 4                         &-4.75        & $\infty$ &0            &$\infty$ &2.094 &2.094 \\
\hline\hline
\end{tabular}
\end{center}
\label{tab1}
\end{table}

\pagebreak

\begin{small}
\begin{figure}
\centerline{\includegraphics[width=160mm, angle=-90]{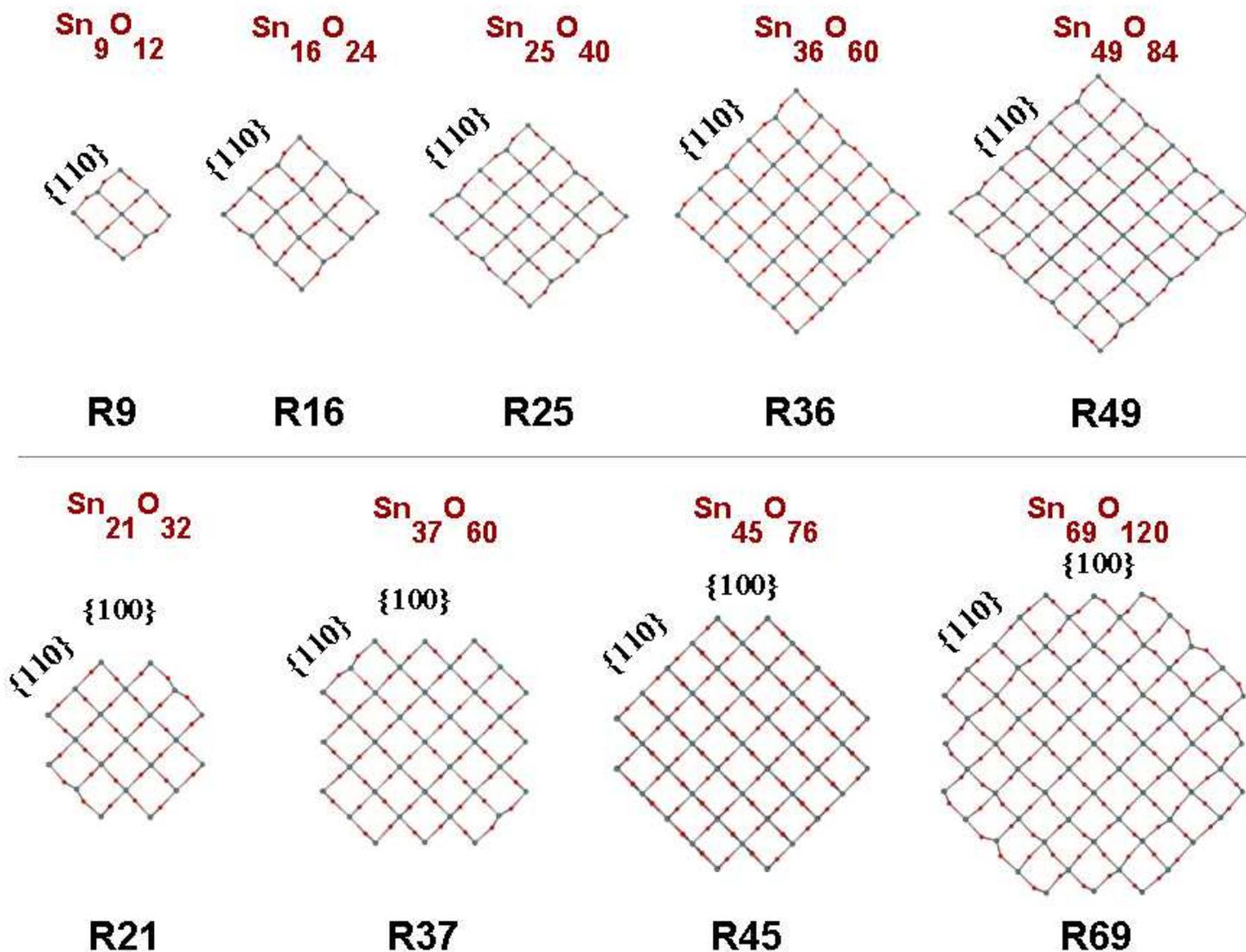}}
\caption{Cross sections of the optimized rutile-like tin oxide nanowires,
grown along the [001] direction. Brown and gray
spheres represent tin and oxygen atoms, respectively.
The wires are grouped in two families, according to types of facet surfaces.
The wires are labeled as R(N), where N is the number of tin atoms
in the unit cell. For family made of pure \{110\} surfaces, wires
were labeled as R9, R16, R25, R36, R49, while
for the family with a mix of \{110\} and \{100\} surfaces, they
were labeled as R21, R37, R45, and R69.}
\label{fig1}
\end{figure}
\end{small}
\pagebreak

\begin{small}
\begin{figure}
\centerline{\includegraphics[width=140mm, angle=270.0]{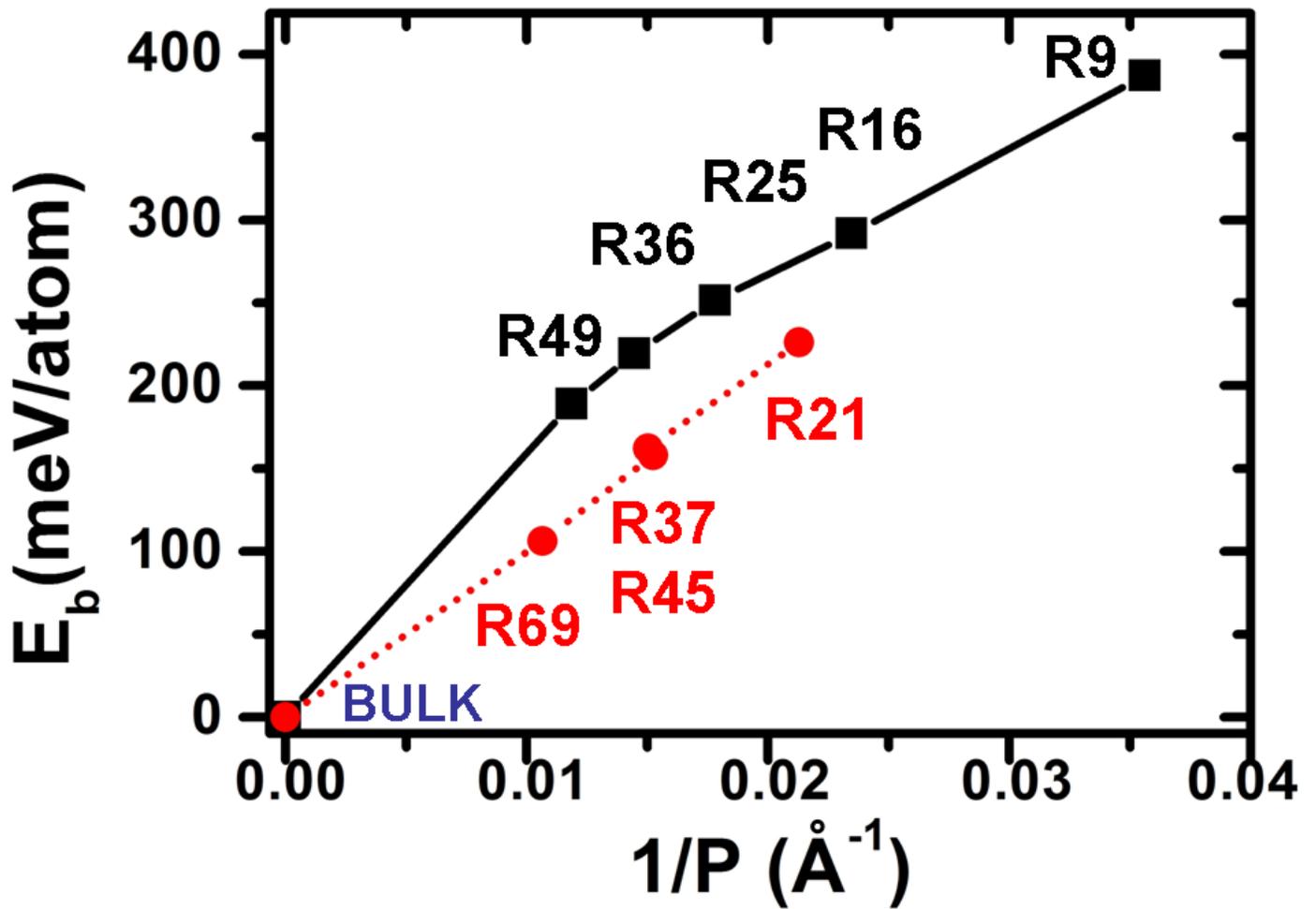}}
\caption{Binding energy ($E_{\rm b}$) (in meV/atom) of the two nanowire
families as a function of the inverse of their perimeters (1/P) (in \AA$^{-1}$). The
energies are given with respect of the respective value
for the bulk SnO$_{2}$ ($E_{b}^{bulk}$ = -4.75 eV).}
\label{fig4}
\end{figure}
\end{small}
\pagebreak

\begin{small}
\begin{figure}
\centerline{\includegraphics[width=150mm, angle=270.0]{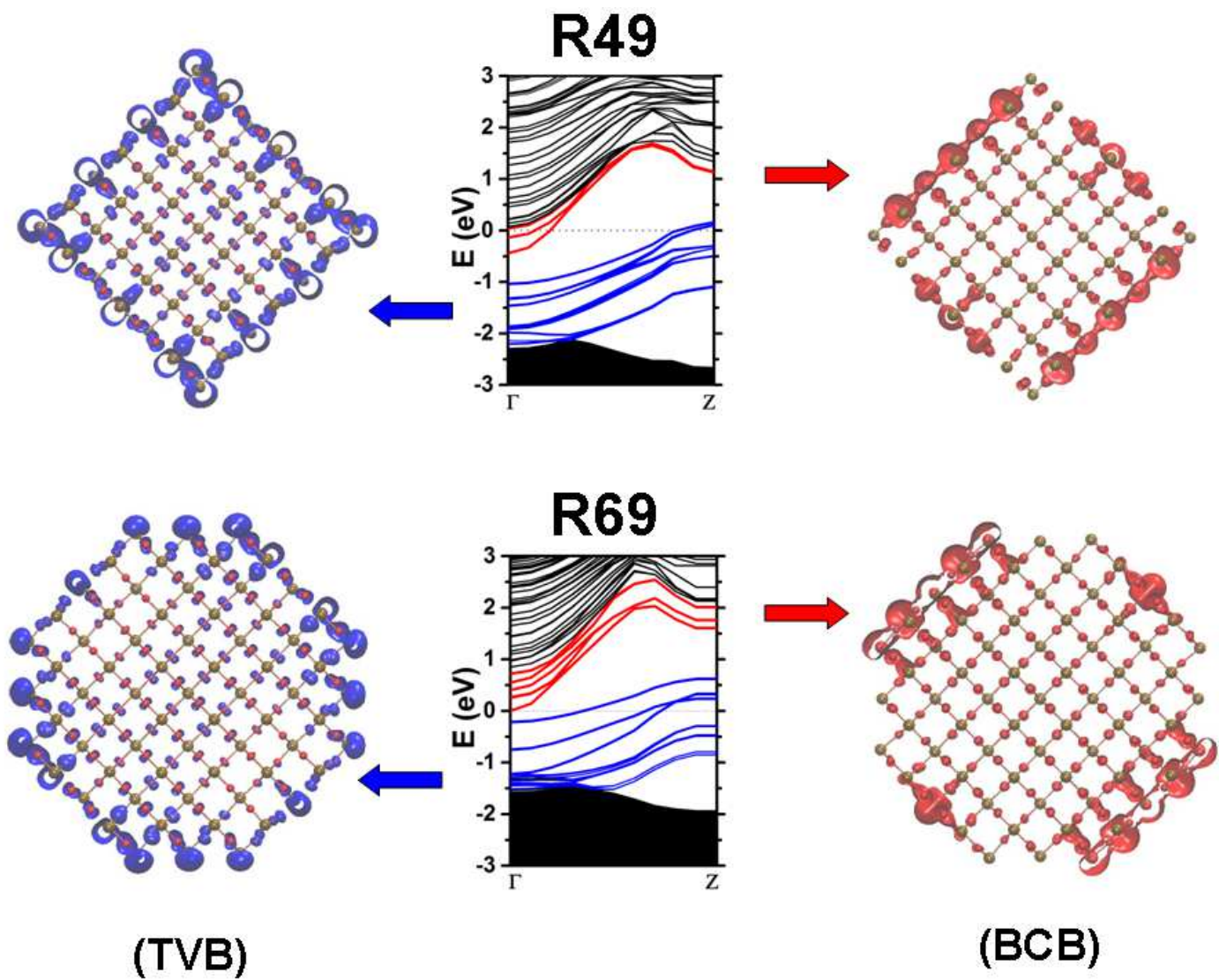}}
\caption{Electronic band structure of nanowires of two different
families (R49 and R69 from \ref{fig1}). The figure shows the probability
density isosurfaces for surface atoms around TVB
(top of the valence band, in blue color) and BCB
(bottom of the conduction ban, in red color) states in
the band sums over 1 {\rm x} 1 {\rm x} 11 Monkhorst{\rm -}Pack special points.
Brown and red spheres represent tin and oxygen atoms, respectively.
Each isosurface corresponds to {\rm 1\%}  of the respective maximum
probability. The shaded regions in the electronic band
structure correspond to bulk states.}
\label{fig2}
\end{figure}
\end{small}
\pagebreak

\begin{small}
\begin{figure}
\centerline{\includegraphics[width=160mm, angle=0.0]{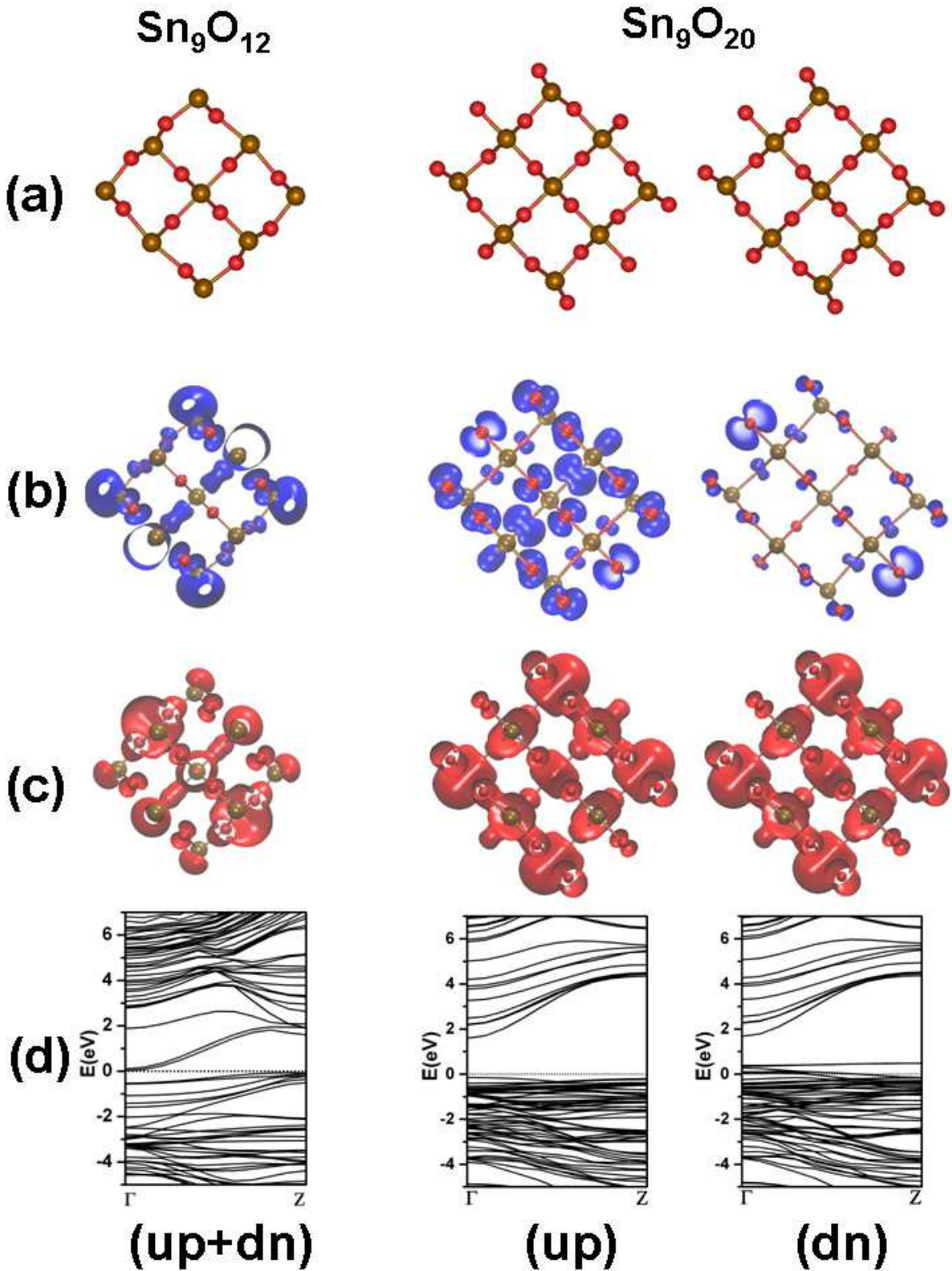}}
\caption{The optimized structure of the unpassivated (Sn$_{9}$O$_{12}$, R9)
and oxygen-passivated (Sn$_{9}$O$_{20}$) nanowires. The figure shows (a) the
relaxed structures, and the probability density isosurfaces for (b) TVB
(blue color) and (c) BCB (red color) states, and (d) the respective band
structures.  Brown and red spheres represent tin and oxygen atoms, respectively.
Each isosurface corresponds to {\rm 1\%} of the respective maximum probability.
For the passivated structure, the results are presented for
spin-up and spin-down electrons. The Fermi energy in (d) is defined at E = 0 eV.}
\label{fig3}
\end{figure}
\end{small}
\pagebreak

\begin{small}
\begin{figure}
\centerline{\includegraphics[width=200mm, angle=0.0]{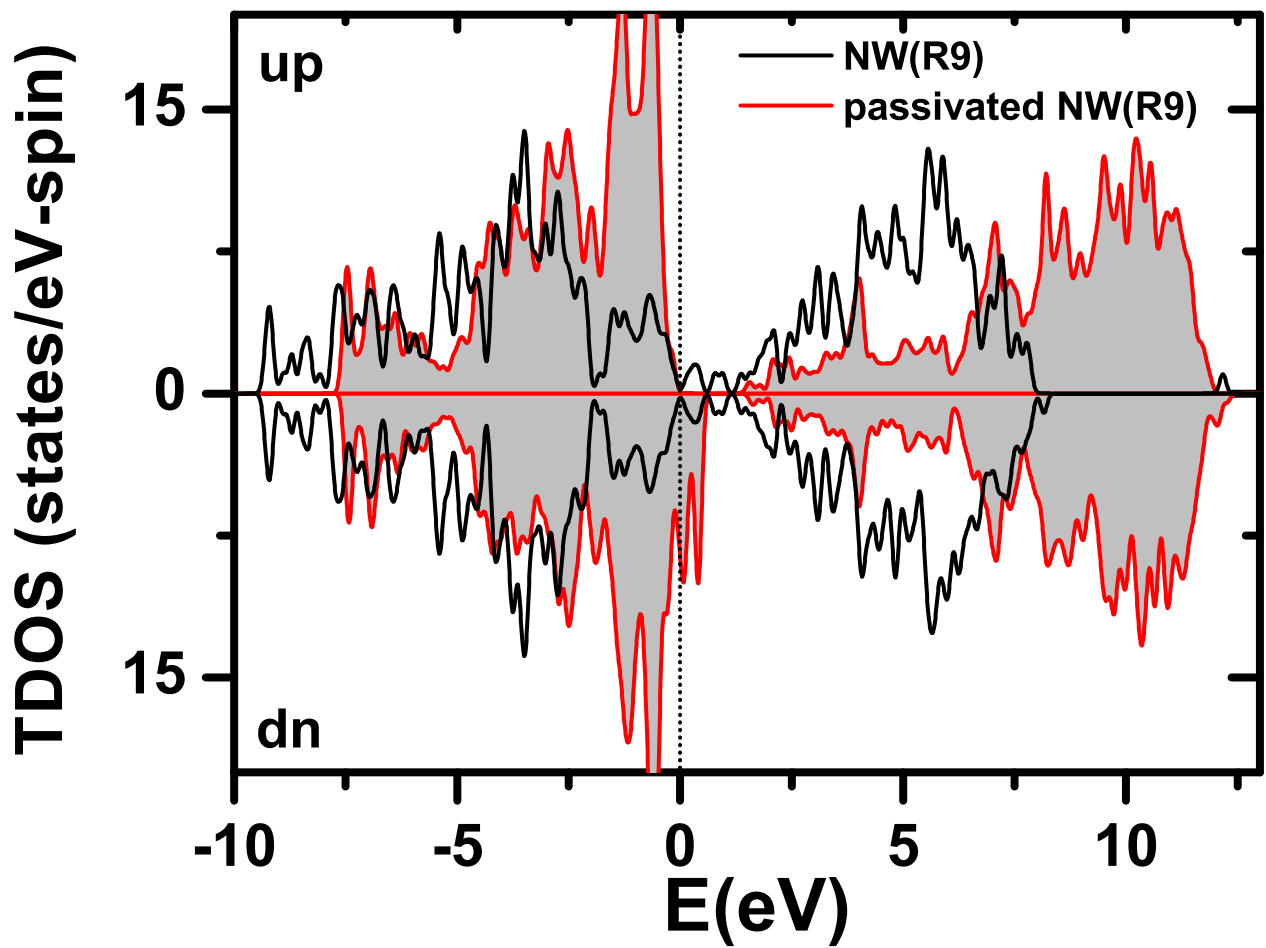}}
\caption{Total electronic density of states (TDOS) for unpassivated and
passivated nanowires (\ref{fig3}a), respectively
Sn$_{9}$O$_{12}$ and Sn$_{9}$O$_{20}$. The figure shows
the contributions from oxygen-passivated (shaded gray region)
and unpassivated (solid black lines) nanowires,
 for the spin-up and spin-down electronic contributions.
The Fermi energy is defined at E = 0 eV.}
\label{fig3a}
\end{figure}
\end{small}
\pagebreak

\end{document}